\documentclass[twocolumn,
aps,prb,twoside,superscriptaddress,floatfix,tightenlines,nofootinbib]{revtex4-1}
\usepackage{dcolumn}
\usepackage{float}
\usepackage{bm}
\usepackage[T1]{fontenc}
\usepackage{amsmath}
\usepackage{graphicx} 
\usepackage{tabularx}
\usepackage{multirow}
\usepackage{graphics}
\usepackage[version=4]{mhchem}
\usepackage[retainplus]{siunitx}
\usepackage{rotating} 
\usepackage{makecell}
\usepackage{hyperref}
\hypersetup{colorlinks=true, linkcolor=blue, citecolor=blue, urlcolor=blue}
\usepackage{tikz}
\usepackage[normalem]{ulem}
\usepackage[utf8]{inputenc}
\usepackage{booktabs}
\usepackage{mathtools}
\usepackage{siunitx}
\usepackage[lofdepth,lotdepth,caption=false]{subfig}
\usepackage[export]{adjustbox}

\begin{document}

\title{Manipulation of spin orientation via ferroelectric switching in Fe-doped Bi$_2$WO$_6$ from first principles}

\author{Katherine Inzani*}
\affiliation{Materials Science Division, Lawrence Berkeley National Laboratory, Berkeley, CA 94720, USA}
\affiliation{Molecular Foundry, Lawrence Berkeley National Laboratory, Berkeley, CA 94720, USA}

\author{Nabaraj Pokhrel*}
\affiliation{Department of Physics, University of California, Merced, CA 95343, USA}

\author{Nima Leclerc}
\affiliation{Molecular Foundry, Lawrence Berkeley National Laboratory, Berkeley, CA 94720, USA}
\affiliation{Department of Electrical and Systems Engineering, University of Pennsylvania, Philadelphia, PA 19104, USA}

\author{Zachary Clemens}
\affiliation{Department of Materials Science and Engineering, University of California, Merced, CA 95343, USA}

\author{Sriram P. Ramkumar}
\affiliation{Department of Materials Science and Engineering, University of California, Merced, CA 95343, USA}







\author{Sin\'{e}ad M. Griffin}
\affiliation{Materials Science Division, Lawrence Berkeley National Laboratory, Berkeley, CA 94720, USA}
\affiliation{Molecular Foundry, Lawrence Berkeley National Laboratory, Berkeley, CA 94720, USA}

\author{Elizabeth A. Nowadnick}
\affiliation{Department of Materials Science and Engineering, University of California, Merced, CA 95343, USA}

\date{\today}

\begin{abstract}
Atomic-scale control of spins by electric fields is highly desirable for future technological applications. Magnetically-doped Aurivillius-phase oxides present one route to achieve this, with magnetic ions substituted into the ferroelectric structure at dilute concentrations, resulting in spin--charge coupling. However, there has been minimal exploration of the ferroelectric switching pathways in this materials class, limiting predictions of the influence of an electric field on magnetic spins in the structure. Here, we determine the ferroelectric switching pathways of the end member of the Aurivillius phase family, \ce{Bi2WO6}, using a combination of group theoretic analysis and density functional theory  calculations. We find that in the ground state $P2_1ab$ phase, a two-step switching pathway via $C2$ and $Cm$ intermediate phases provides the lowest energy barrier. Considering iron substitutions  on the W-site in \ce{Bi2WO6}, we determine the spin easy axis. By tracking  the change in spin directionality during ferroelectric switching, we find that a 90$^\circ$ switch in the polarization direction leads to a 112$^\circ$ reorientation of the spin easy axis. The low symmetry crystal-field environment of Bi$_2$WO$_6$ and magnetoelastic coupling on the magnetic dopant provide a route to spin control via an applied electric field.
\end{abstract}

\maketitle

\section{Introduction}
Multiferroic materials with coupled ferroic orderings (e.g. ferromagnetism, ferroelectricity, ferroelasticity) 
exhibit intriguing physics and hold potential for enabling new types of  future electronic devices.~\cite{Spaldin2019} In magnetoelectric materials with coupled ferroelectricity  and magnetism, the ability to switch the magnetization by an applied electric field is particularly promising for low-power spintronics.~\cite{Manipatruni2019} Materials approaches to multiferroicity, for example multiferroic superlattices~\cite{Mundy2016}, nanocomposites~\cite{Cai2017}, domain walls~\cite{Catalan2012} and single phase materials~\cite{Wang2003c}, typically focus on realizing long-range magnetic order for macroscopic devices. However, several recent works have pushed towards 
the fundamental limits of multiferroic phenomena, including electric field manipulation of molecular magnets~\cite{Boudalis2018,Liu2019e}, tuning exchange in a molecular system~\cite{Fittipaldi2019}, and the coherent electric field control of dilute iron dopants in a ferroelectric crystal.~\cite{Liu2021} These milestones towards full control of isolated spins by electric fields may enable new functionalities in  
classical electronic devices in the field of spintronics as well as in quantum computing.~\cite{Johnson2019,Liu2021}

A promising pathway to achieve isolated spin centers with magnetoelectric coupling is to dope a ferroelectric structure with dilute concentrations of magnetic ions.~\cite{Liu2021} Here, magnetoelectricity arises by coupling the ferroelectric's polar distortion with the spin dopant through spin-orbit interactions. This approach confers the rich phase space of ferroelectric crystals, in particular of complex oxide materials, for use as hosts for spin dopants. In particular, the versatile  structural motifs and distortions in ferroelectric oxides provide a highly tunable local environment for the spin center, allowing control of the magnetocrystalline properties via the crystal field environment. The symmetry lowering caused by the ferroelectric distortion results in magnetocrystalline anisotropies that  lead to preferential alignment of spins within a plane (spin easy plane) or along an axis (spin easy axis). Using the prototypical ferroelectric \ce{PbTiO3} as a host for Fe$^{3+}$ spins, some of the present authors recently demonstrated that the tetragonal polar distortion results in a spin easy plane with 90$\degree$ switching under the application of an electric field.~\cite{Liu2021} However, 
preferentially aligning spins along an easy
axis and 180$\degree$ switching would have technical advantages for  applications. Ferroelectric hosts providing lower crystallographic symmetries in the vicinity of the spin are more likely to support spin easy axes due to their highly distorted crystal fields, and so are sought for such ferroelectric-mediated spin switching.

The Aurivillius phases are a family of layered ferroelectric materials with low symmetry crystal structures that could satisfy these requirements. The Aurivillius structure is composed of $m$ perovskite-like layers ($A_{m-1}B_m$O$_{3m+1}$)$^{2-}$ interspersed with fluorite-like (\ce{Bi2O2})$^{2+}$ layers, giving the overall general formula Bi$_2A_{m-1}B_m$O$_{3m+3}$. The Aurivillius phases are well known for their 
robust ferroelectricity, including high Curie temperatures (T\textsubscript{c}), large spontaneous polarizations~\cite{Utkin1980,Galasso2007,Campanini2019}, and fatigue resistance.~\cite{de1995fatigue} Furthermore, the composition has great versatility owing to the different cations that can be placed on the $A$ and $B$ sites, which has led to efforts to design multiferroic
 Aurivillius compounds
 via incorporation of magnetic ions.~\cite{Mao2009,Keeney2013,Chen2014a,Wang2015c,Li2016a,Keeney2012,Kalani_et_al:2021} Most work has focused on achieving long range magnetic ordering in single phase materials with large proportions of magnetic cations, 
 for example in doped Bi$_{n+1}$Fe$_{n-3}$Ti$_3$O$_{3n+3}$ compounds. However, the complex crystal structure and difficulty in synthesis of phase pure samples has made  
 characterization of the multiferroic properties difficult.~\cite{Keeney2012,Birenbaum2014,Zhai2018} 
 In particular, the ferroelectric and magnetoelectric switching mechanisms have not been elucidated.~\cite{Faraz2017}
These limitations hinder the prediction of the behavior of magnetic spins during switching. Moreover, to the best of our knowledge, magnetoelectric coupling of isolated magnetic dopants has not yet been investigated in this class of materials. 

Here, we use group theoretic analysis and first principles calculations to explore ferroelectric switching and control of magnetic dopants in the end member of the Aurivillius family, \ce{Bi2WO6} ($m$=1, $B$=W). We select \ce{Bi2WO6} because it exhibits robust ferroelectricity and also possesses a complex crystal structure which can provide a low symmetry crystallographic environment for magnetic dopants. Theoretical and experimental work has revealed that ferroelectricity in \ce{Bi2WO6} arises from an instability to a polar distortion involving large Bi displacements with respect to the perovskite layer.~\cite{Machado2004,Mohn2011,djani2012first,Wang2016d,Okudera2018} It undergoes a two-step paraelectric-ferroelectric phase transition sequence:  at room temperature, \ce{Bi2WO6} crystallizes in the orthorhombic ferroelectric phase $P2_1ab$, then  transitions to the polar orthorhombic structure $B2cb$ at  670$\degree$C, and  finally  transitions to the paraelectric monoclinic phase $A2/m$ above 950$\degree$C.~\cite{BWO_expt1}

Experiments have reported that ferroelectric switching in \ce{Bi2WO6} proceeds via a two-step process~\cite{Wang2016d}, but the details of the precise switching pathway taken are still lacking. We therefore start by determining the likely  ferroelectric switching pathway, by systematically enumerating and then evaluating the energetics of several possible symmetry-distinct paths. Here we consider \textit{intrinsic} ferroelectric switching paths, where we calculate energy barriers for  coherent polarization reversal in a single infinite domain.~\cite{beckman2009ideal} Although this does not provide a full description of the dynamic ferroelectric switching process, work on other ferroelectrics~\cite{heron2014deterministic,nowadnick2016domains} has shown that when multiple symmetry-distinct switching paths are available,  intrinsic barriers  can correctly identify the experimental switching  path.  We then introduce Fe$^{3+}$ dopants into the structure at dilute concentrations and track the change in spin directionality with ferroelectric switching. This work lends understanding to the magnetoelectric effects on isolated spins in \ce{Bi2WO6}, demonstrating the potential for atomic-scale spin control in this class of materials.

\section{Computational Methodology}
\noindent
We perform density functional theory (DFT) calculations  using the Vienna \textit{Ab initio} Simulation Package (VASP)~\cite{Kresse1993,Kresse1994,Kresse1996,Kresse1996a}, using projector augmented wave (PAW) pseudopotentials~\cite{Blo,Kresse1999} including Bi $5d^{10}6s^{2}6p^{3}$, W $5p^{6}5d^{4}6s^{2}$, O $2s^{2}2p^{4}$ and Fe $3d^{6}4s^{2}$ as valence electrons.
A plane wave cut-off energy of 800~eV is used with a $6\times6\times2$ Gamma-centered \textit{k}-point grid (for the $P2_1ab$ \ce{Bi2WO6} 36-atom unit cell), which converges the total energy to 1~meV per formula unit (f.u.). The same $k$-point grid is used for all undoped structures, and a $2\times2\times2$ \textit{k}-point grid is used for the doped supercells. All calculations are done using the generalized gradient approximation (GGA) based exchange-correlation functional PBEsol\cite{Perdew2008}, which gives lattice parameters within 1\% of experiment~\cite{Okudera2018} (\textit{a}=5.443~\AA, \textit{b}=5.443~\AA, \textit{c}=16.557~\AA, for $P2_1ab$).

For undoped structures, we allow the ionic positions, cell volume, and cell shape to optimize and apply a force tolerance convergence of \SI{1}{\milli\electronvolt\per\angstrom}. The nudged elastic band (NEB) method~\cite{NEB} implemented in VASP is employed to find the structural parameters and energies of intermediate structures lying along the ferroelectric switching pathways. For the NEB calculations, the force convergence tolerance is increased to \SI{10}{\milli\electronvolt\per\angstrom}. 

We calculated Fe$^{3+}$ substitutional defects on W-sites in $2\times2\times1$ supercells of \ce{Bi2WO6} (144 atoms), with three electrons added for charge compensation. For structures containing more than one symmetrically inequivalent W-site, we consider each as a different dopant site. An effective Hubbard term $U$\textsubscript{eff} = $U$–$J$ = 4~eV is added to the Fe \textit{d}-orbitals within the Dudarev approach.~\cite{Dudarev1998} Geometry optimization of the ions for each defect supercell is completed to a force convergence of \SI{10}{\milli\electronvolt\per\angstrom}, whilst keeping the cell volume and shape fixed. Magnetocrystalline anisotropy energy (MCAE) surfaces are calculated by including spin-orbit coupling self consistently and varying the spin quantization axes over 194 points. We make use of the ISOTROPY software suite~\cite{isotropy} for group theoretic analysis and VESTA~\cite{vesta} for the visualization of crystal structures.

\section{Results and discussion}
\noindent

\subsection{Ground state crystal structure}
\noindent


\begin{figure*}
 \centering
 \includegraphics[width=0.9\linewidth]{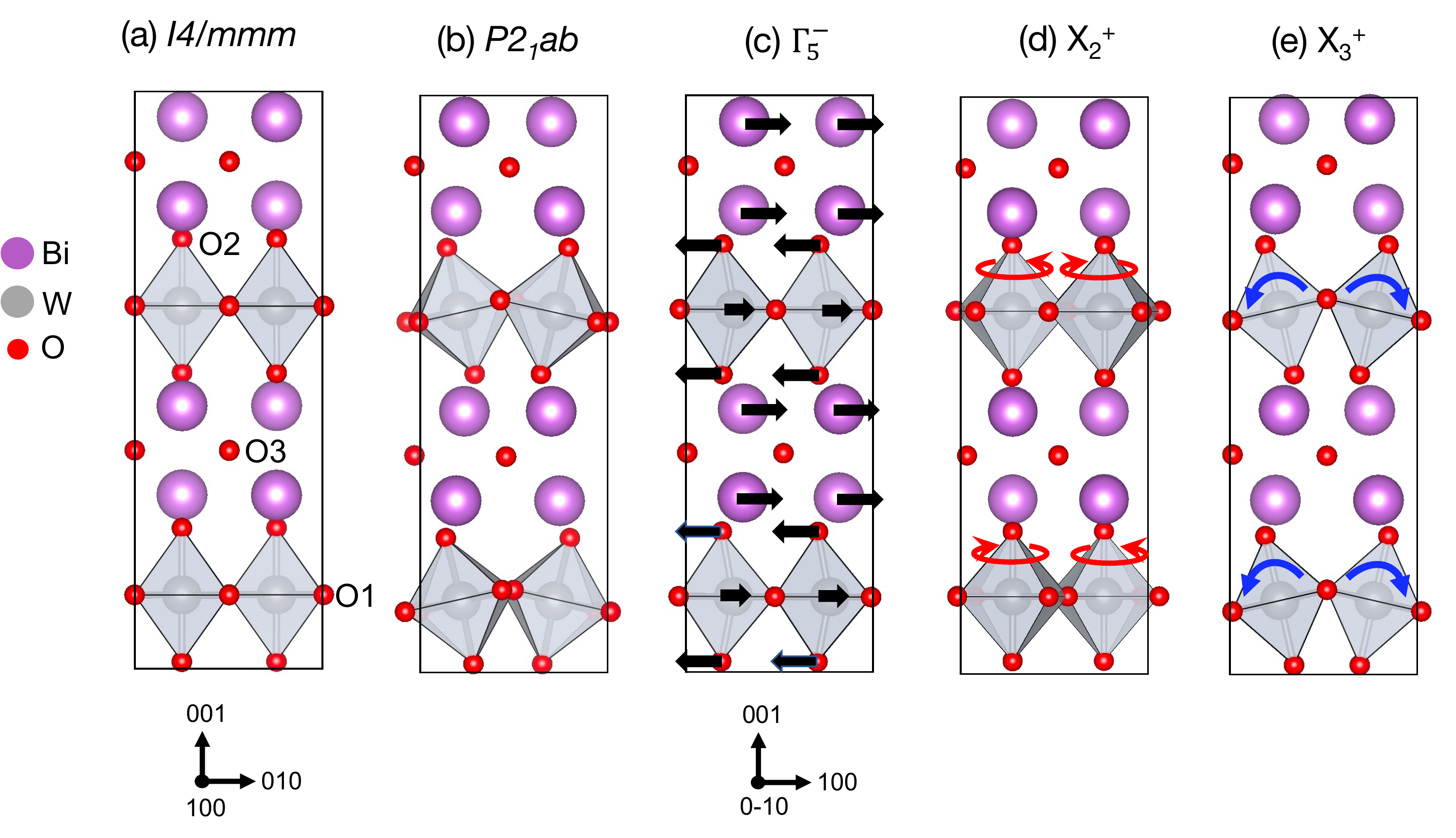}
 \captionsetup{width=0.75\linewidth,justification=raggedright}
 \caption{ The (a) high-symmetry reference structure $I4/mmm$ and (b) polar orthorhombic structure $P2_1ab$ of   Bi$_2$WO$_6$.  A structural decomposition of $P2_1ab$ into symmetry adapted modes of $I4/mmm$ reveals three main  structural distortions: (c)  a polar displacement along [1~0~0] with symmetry $\Gamma_5^-$, (d) an octahedral rotation about [0~0~1] with symmetry $X_2^+$, and (e) an out-of-phase octahedral tilt  about [1~0~0] with symmetry $X_3^+$. The axes shown under panel (a) are used for all panels except for (c).
}
 \label{fig:distortions}
\end{figure*}

To set the stage for understanding ferroelectric switching, we first analyze the structural distortions present in the ferroelectric $P2_1ab$ structure.
The   $P2_1ab$ space group is established by the condensation of three distinct structural distortions that transform like irreducible representations (irreps) of the high-symmetry reference  structure $I4/mmm$ (Fig.~\ref{fig:distortions}a, b).  These distortions are a polar displacement along the [1~0~0] orthorhombic axis which transforms like the irrep $\Gamma_5^-$, an  octahedral rotation  about [0~0~1]  which transforms like $X_2^+$, and  an out-of-phase ($a^-a^-c^0$ in Glazer notation~\cite{glazer1972}) octahedral tilt about   [1~0~0]   which transforms like  $X_3^+$,  (Fig.~\ref{fig:distortions}c-e).  
The amplitudes of these three distortions, computed from   DFT-relaxed and experimental structures, are reported in  Table \ref{tab:decomp}. Overall the distortion amplitudes show good agreement between DFT and experiment. The main contribution to the polar distortion comes from displacement of the Bi cations against the O2 atoms as shown  in  Fig. \ref{fig:distortions}(c). In addition to the three distortions discussed above, there are several other distortions which are symmetry-allowed  in the $P2_1ab$ structure, but they have negligible amplitudes~\cite{djani2012first} so we do not consider them in this work. 



\begin{table}[]
\caption{\label{tab:decomp} Decomposition of the Bi$_2$WO$_6$ $P2_1ab$ structure (DFT-relaxed and experimental) into symmetry adapted modes of $I4/mmm$. The experimental structure is taken from Ref. {\onlinecite{Rae:al0453}}. The  amplitudes are given in \AA~ for a 36-atom computational cell. For the O1 atoms, the coordinates in parentheses indicate the axes along which the atoms displace; there are two distinct O1 displacement patterns in the $xy$ plane that are consistent with the symmetry.  
}
\begin{tabular}{lccccccccc}
\hline
\hline
Atom    & \multicolumn{2}{c}{$\Gamma_5^-$} && \multicolumn{2}{c}{$X_2^+$} && \multicolumn{2}{c}{$X_3^+$} \\
         &  DFT & Expt.     &&  DFT & Expt. &&  DFT & Expt.\\
         \hline
Bi           & 0.66 &0.68 &&   0 &0        && 0.33 &0.30\\
W            & 0.18 &0.25  && 0 &0        && 0 &0\\
O1($z$) & 0 &0        &&  0 &0       && 0.87 &0.70\\
O1($xy$) & -0.15 &-0.11 && 0.75 &0.85  && 0 & 0\\
O1($xy$) & -0.25 &-0.29 && -0.01 &-0.02 && 0 &0\\
O2           & -0.77 &-0.74 && 0 &0        && -1.10 &-0.89\\
O3           & 0.27 &0.17  && 0 &0        && -0.05 &-0.02\\ 
\hline
Total & 1.10 &1.09 &&   0.75 &0.85        && 1.44 &1.17\\
                      \hline
                      \hline
\end{tabular}
\end{table}


\begin{table*}
\caption{\label{tab:irreps_details} Subgroups of $I4/mmm$ established by different combinations of the $X_3^+$, $X_2^+$, and $\Gamma_5^-$ order parameter directions. Total energies, distortion amplitudes, and lattice parameters obtained from DFT structural relaxations of Bi$_2$WO$_6$ in each space group are given. The energies are reported relative to the energy of $P2_1ab$, which is set to 0 meV/f.u.. If a structure relaxes to a higher symmetry space group, that space group is indicated in the energy column. The distortion amplitudes  are obtained by decomposing the distorted structures with respect to $I4/mmm$ and are reported for a 36-atom computational cell. }

\begin{tabular}{c  c  c  c  c c c c c  c c c c c }
\hline
\hline
Irreps & \multicolumn{3}{l}{Order parameter direction} && Space && \multicolumn{3}{l}{Amplitude ({\AA}})& \multicolumn{3}{l}{Lattice parameters ({\AA}}) & Energy  \\
       &   $\Gamma_5^-$ & ${X_2}^+$ &  ${X_3}^+$   && group (N$^{o}$)  && $\Gamma_5^-$ & $X_2^+$ & $X_3^+$ & $a$ & $b$ & $c$ &(meV/f.u.)\\
\\
\hline
-                &   -     & - & - && $I4/mmm$ (139) && 0 & 0 & 0 &  3.806 & 3.806 & 16.453 &355.51\\
\hline
                                    
$X_2^+ \oplus X_3^+$             & - & $(b,0)$ & $(c,0)$ && $P2_1/c$                                 (14)  &&  0   & 0.91   & 1.30  &   5.331  &   5.353 &   8.697   & 158.72 \\   

                                    &   -     & $(b,0)$ & $(0,c)$ && $Pcab$ (61) && 0 & 0.78 & 1.35 &  5.350 & 5.360 & 16.844 &149.74\\
                                    &  -      & $(b,b)$ & $(c,c)$ && $C2/m$ (12) && 0 & 0.48 &  1.46   &  7.602  & 7.614 & 16.602 & 157.40\\      
                                    \hline
$\Gamma_5^- \oplus X_3^+$           & $(a,0)$ & - & $(c, c)$ && $Cm2a$ (39) &&  0.44 &  0 & 1.53 &  7.601 &  7.754 & 16.481   & 103.46 \\
                                    & $(a,a)$ & - & $(0,c)$ && $B2cb$ (41)  && 1.19    &  0  &  1.60   &  5.467       & 5.472     &  16.553      & 4.38\\
                                    
                                     & $(a,a)$ & - & $(c,0)$ && $Bb2_1m$ (36)  &&   &    &   &  &   &   & $(Fmm2)$\\
\hline
$\Gamma_5^- \oplus X_2^+$           & $(a,a)$ & $(b, 0)$ & - && $Bb2_1m$ (36) &&   0.98   &  1.24    &   0   &  5.389     & 5.395           &  16.481  & 67.05 \\

                                    & $(a,a)$ & $(0,b)$ & - && $Cm2a$ (39) && 1.01 & 1.22 & 0  &  7.601    &  7.754 & 16.481    & 71.21\\
                                    
                                    & $(a,0)$ & $(b,b)$ & - && $Cm2m$ (38) &&  1.15  & 0.74   &   0   &  7.544     & 7.845 &    16.569    & 133.20\\
\hline
$\Gamma_5^- \oplus X_2^+ \oplus X_3^+$ & $(0, a)$ & $(b, b)$ & $(c, c)$ && $Cm$ (8) && 0.45 & 0.53 &  1.46 & 7.729 & 7.584 & 16.506 & 99.06\\

                                      & $(a,0)$ & $(b, b)$ & $(c, c)$ && $C2$ (5) && 0.49 & 0.58 &  1.45 &
7.570 &  7.740 & 16.507 & 96.68\\

                                       & $(a,a)$ & $(b,0)$  & $(c,0)$  &&  $P2_1$ (4) &&   &     &      &  
      &    &

      &  ($Bb2_1m$) \\
   
                                        & $(a,a)$ & $(0,b)$ & $(0,c)$ &&  $Pc$ (7) && 1.11 & 0.76 & 1.44 
      &        5.442 & 5.443        & 8.688     & -0.08 \\
                                       & $(a,a)$ & $(b,0)$ & $(0,c)$ &&  $P2_1ab$ (29) && 1.11  & 0.75 & 1.44 
      &        5.443 & 5.443        & 16.557      & 0 \\
      
    
\hline
\hline
\label{Tab:intermediate_structures}
\end{tabular}
\end{table*}

\subsection{Ferroelectric switching pathways}
\noindent

To provide a framework for systematically identifying ferroelectric switching pathways, we next enumerate possible metastable structural phases of Bi$_2$WO$_6$ and compute their energies with DFT. The key to uncovering the metastable structural phases is to recognize that each of the three structural distortions shown in Figure~\ref{fig:distortions} is described by a two-dimensional order parameter $Qe^{i\alpha}$.~\cite{nowadnick2016domains} Here $Q$ is the order parameter amplitude and $\alpha$ is the phase. For the $X_3^+$ octahedral tilt and the $\Gamma_5^-$ polar distortion, the phase $\alpha$ describes the orientation of the tilt (polar) axis. For the $X_2^+$ octahedral rotation, the phase describes the relative ``sense'' of the octahedral rotations in adjacent perovskite  layers. 

Each two-dimensional structural order parameter can lie along three symmetry-distinct directions ($Q\cos\alpha$, $Q\sin\alpha$)= ($a$,0), ($a$,$a$), or ($a$,$b$) where $a\ne b$ are real numbers and  each direction establishes a different subgroup of $I4/mmm$. For example, the $P2_1ab$ space group is established by condensing the $X_3^+$, $X_2^+$, and $\Gamma_5^-$ distortions along the (0,$a$), ($a$,0) and ($a$,$a$) order parameter directions, respectively. Taking these order parameters  to lie along  different combinations of directions generates  structures of  different symmetries. In Table~\ref{Tab:intermediate_structures} we consider all other possible combinations of the  $X_3^+$, $X_2^+$, and $\Gamma_5^-$ order parameter directions, and enumerate the space groups that these generate (we do not include the low symmetry ($a$,$b$) direction in this enumeration because structures defined by this direction generally return to a higher symmetry direction upon DFT relaxation). 
 Table~\ref{Tab:intermediate_structures} also shows space groups that are generated by combining two out of the three order parameters taken along different combinations of directions. Space groups generated by each order parameter individually are given in Appendix~\ref{appendix1}. 
 
 We then perform structural relaxations  of Bi$_2$WO$_6$ with its symmetry constrained to each space group identified in Table~\ref{Tab:intermediate_structures}, and report the resulting energy and structural parameters. Most metastable structures have energies ranging from $\approx$65 to $\approx$160 meV per f.u. above $P2_1ab$. We find two very low energy structures: $B2cb$ (4.38 meV/f.u.) and $Pc$ which we find to be  slightly lower in energy than $P2_1ab$. The $Pc$ and $P2_1ab$ structures exhibit the same $X_3^+$ and $\Gamma_5^-$ distortions, the only difference is the relative ``sense" of the $X_2^+$ rotations in adjacent perovskite layers, thus it is unsurprising that these phases are very close in energy. We note that the relative energy of $Pc$ and $P2_1ab$ is quite sensitive to the value of the lattice parameters. For example,   Ref.~\onlinecite{djani2012first} found $Pc$ to be about 3 meV/f.u. higher in energy than $P2_1ab$ from DFT calculations with the LDA functional. 
 Since $P2_1ab$ is the experimentally reported ground state, we do not further consider the $Pc$ phase here.
 
\begin{figure}[h!]
 \centering
 \includegraphics[width=1\linewidth]{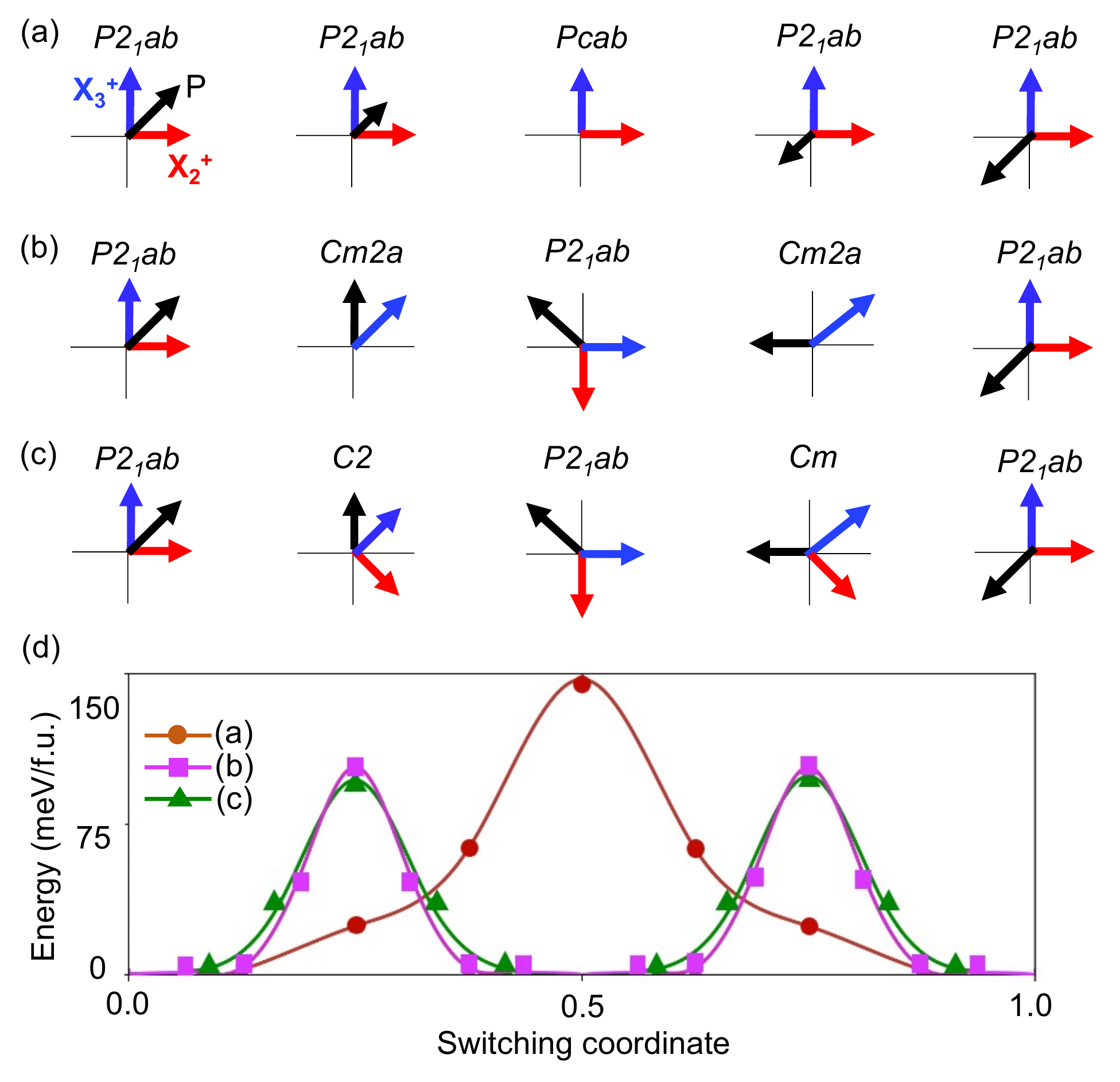}
 \captionsetup{width=1\linewidth,justification=raggedright}
 \caption{Ferroelectric switching pathways in \ce{Bi2WO6}. (a)-(c) show how the $X_3^+$, $X_2^+$, and $\Gamma_5^-$ (\textbf{\textit{P}}) order parameters, denoted by blue, red, and black arrows, respectively, evolve along the switching path. Path (a) is a one-step path, whereas paths (b)-(c) are two-step paths which pass through an orthorhombic twin of $P2_1ab$ at the midpoint of the switching path. (d) Energy as a function of switching coordinate for the paths shown in (a)-(c). 
 }
 \label{P_switching_pathways}
\end{figure}

\begin{figure}[h!]
 \centering
 \includegraphics[width=1\linewidth]{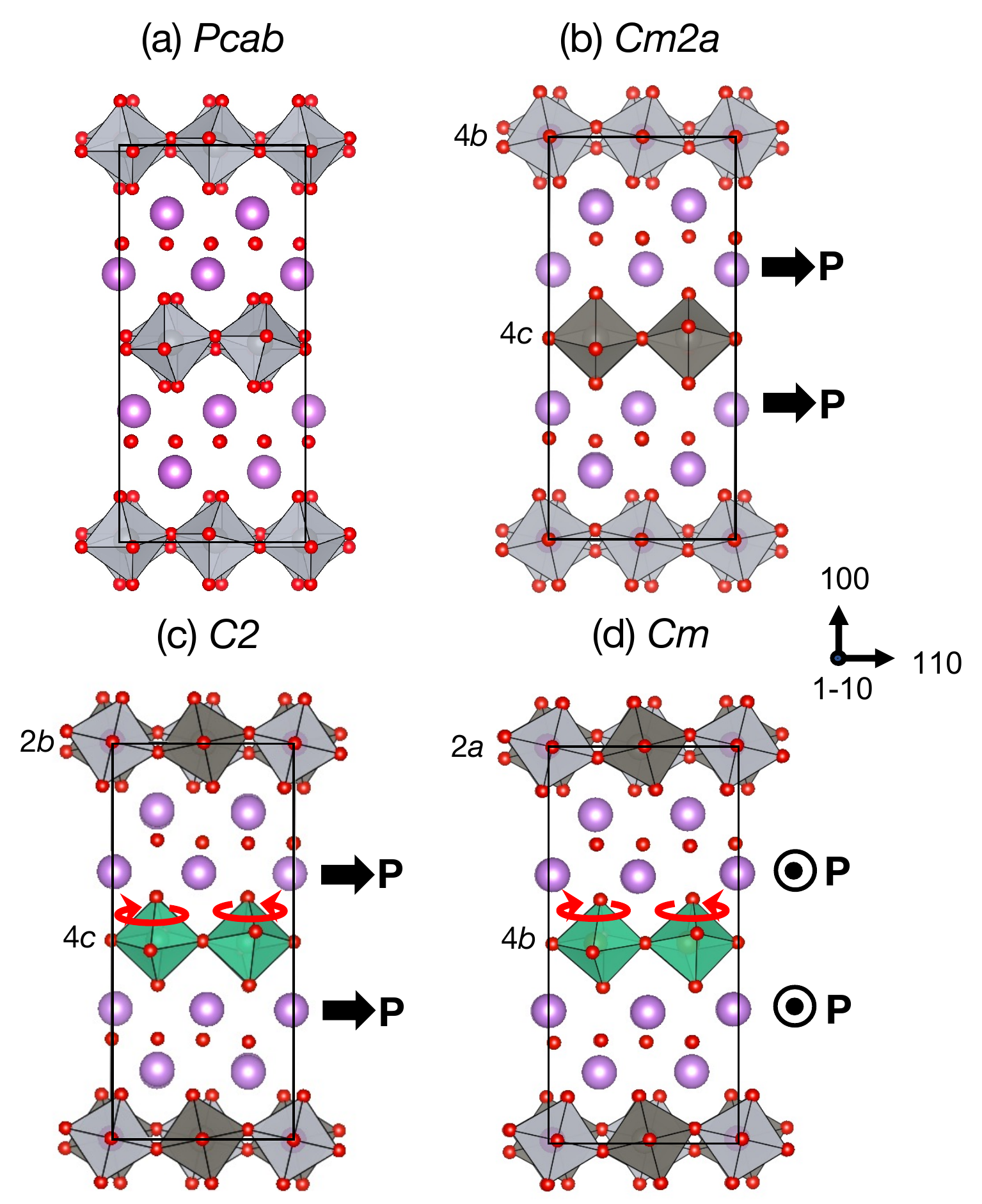}
 \captionsetup{width=0.95\linewidth,justification=raggedright}
 \caption{Barrier structures of Bi$_2$WO$_6$ realized along the ferroelectric switching paths. The $Cm2a$ structure (b) has two distinct W Wyckoff positions, indicated by the light grey (4$b$) and dark grey (4$c$) octahedra. The $C2$ structure (c) also has two distinct W Wyckoff positions, indicated by grey (2$b$) and green (4$c$) octahedra, and two symmetrically inequivalent W-sites on the 2$b$ position indicated in light and dark grey, respectively. Similarly, the $Cm$ structure (d) has two distinct W Wyckoff positions, indicated by grey (2$a$) and green (4$b$) octahedra, and two symmetrically inequivalent W-sites on the 2$a$ position indicated in light  and dark grey, respectively.
 }
 \label{fig:intermediate_structures}
\end{figure}

 We next use the results of Table~\ref{Tab:intermediate_structures} to enumerate possible Bi$_2$WO$_6$ ferroelectric switching pathways.  The simplest way to reverse the polarization is in a single 180$^\circ$ step, where the polarization is brought to zero and then turned on again pointing in the opposite direction, as shown in Figure~\ref{P_switching_pathways}a. At the midpoint of the path,  the amplitude of the polarization is zero, and the symmetry of the crystal structure is $Pcab$, which is 149.74 meV/f.u. above the $P2_1ab$ ground state structure (see Table~\ref{Tab:intermediate_structures}). The $Pcab$ crystal structure  is shown in Figure~\ref{fig:intermediate_structures}a.



In addition to the one-step $Pcab$ switching path, we identify three  ``two-step'' switching pathways, where the polarization reverses direction by rotating through  two 90$\degree$ steps (while maintaining finite amplitude). Since the $\Gamma_5^-$ order parameter in $P2_1ab$ is oriented along the ($a$,$a$) direction, rotating it by 90$^\circ$ takes it to either the ($-a$,$a$) or ($a$,$-a$) direction. This rotation requires that the $\Gamma_5^-$ order parameter pass through the (0,$a$) or ($a$,0) direction.  Table~\ref{Tab:intermediate_structures} reveals that the $Cm2a$, $Cm2m$, $C2$, and $Cm$ structures satisfy this requirement. Interestingly, the energies of $Cm2a$, $C2$, and $Cm$ are all near 100 meV/f.u. (within 10 meV/f.u. of each other), whereas $Cm2m$ is somewhat higher (133.20 meV/f.u.). Due to its higher barrier, we do not consider the $Cm2m$ pathway further in this work.

Using these identified structures, we construct the two lowest energy two-step ferroelectric switching pathways in  Figure~\ref{P_switching_pathways}(b-c).  Figure~\ref{P_switching_pathways}(b) shows a pathway that passes through $Cm2a$  twice as the polarization rotates in two 90$^\circ$ steps. At the midpoint of the switching path, the structure passes through an  orthorhombic twin domain of $P2_1ab$. Note that the structure passes through different domains of $Cm2a$ in the first and second steps. As the polarization rotates counterclockwise, the $X_3^+$ order parameter  rotates clockwise by 90$^\circ$ in the first step, and then rotates back to its original orientation in the second step. In each step, the $X_2^+$ order parameter turns off so that it reaches zero at the $Cm2a$ structure, and then turns on again reoriented by 90$^\circ$. The $Cm2a$ structure is shown in Figure~\ref{fig:intermediate_structures}(b). Here the ($a$,$a$) direction of $X_3^+$ establishes an octahedral tilt pattern where the tilt axes of adjacent perovskite layers are perpendicular to each other, so that there are $a^-b^0b^0$ and $b^0a^-b^0$ rotations in the dark and light grey perovskite layers in Figure~\ref{fig:intermediate_structures}(b), respectively. Note that the higher energy $Cm2m$ pathway follows a similar evolution of structural order parameters as the $Cm2a$ path, except the $X_2^+$ rather than the $X_3^+$ order parameter rotates during the switching process.

The second two-step switching path that we investigate is shown in Figure~\ref{P_switching_pathways}(c). The $\Gamma_5^-$ and $X_3^+$ order parameters follow the same sequence as in Figure~\ref{P_switching_pathways}(b), except now the $X_2^+$ order parameter makes two 90$^\circ$ rotations while maintaining finite amplitude, rather than turning off/on. The barrier structure in the first step has symmetry $C2$, whereas in the second step it has symmetry $Cm$. The $C2$ and $Cm$ structures are shown in Figure~\ref{fig:intermediate_structures}(c) and (d), respectively. These structures have the same $a^-b^0b^0/b^0a^-b^0$ octahedral tilt pattern as $Cm2a$. The $C2$ and $Cm$ structures share the same $X_2^+$ rotation pattern, where  every other perovskite layer (those colored green in Figure~\ref{fig:intermediate_structures}(c-d)) exhibit finite amplitude rotations about [0~0~1], and the other (grey) layers have no $X_2^+$ rotation amplitude. The difference between the $C2$ and $Cm$ structures is the orientation of the polarization with respect to the $X_3^+$ and $X_2^+$ order parameters: in $C2$ the polarization lies along the $X_3^+$ tilt axis of the green octahedra which have finite $X_2^+$ rotations, whereas in $Cm$  the polarization lies along the $X_3^+$ tilt axis of the grey octahedra which have no $X_2^+$ rotation. 

In order to investigate how the energy changes during switching,   Figure~\ref{P_switching_pathways}(d) shows nudged elastic band (NEB) calculations of the energy as a function of 
 switching coordinate for the paths in Figure~\ref{P_switching_pathways}(a-c). Both two-step paths have a significantly lower energy barrier than the one-step $Pcab$ path, with the energy barriers for the $C2/Cm$ path (96.68 and 99.06 meV/f.u. in the first and second steps, taken from Table~\ref{Tab:intermediate_structures}) being slightly lower than the $Cm2a$ path (103.46 meV/f.u.). 

\begin{figure}[h!]
 \centering
 \includegraphics[width=1\linewidth]{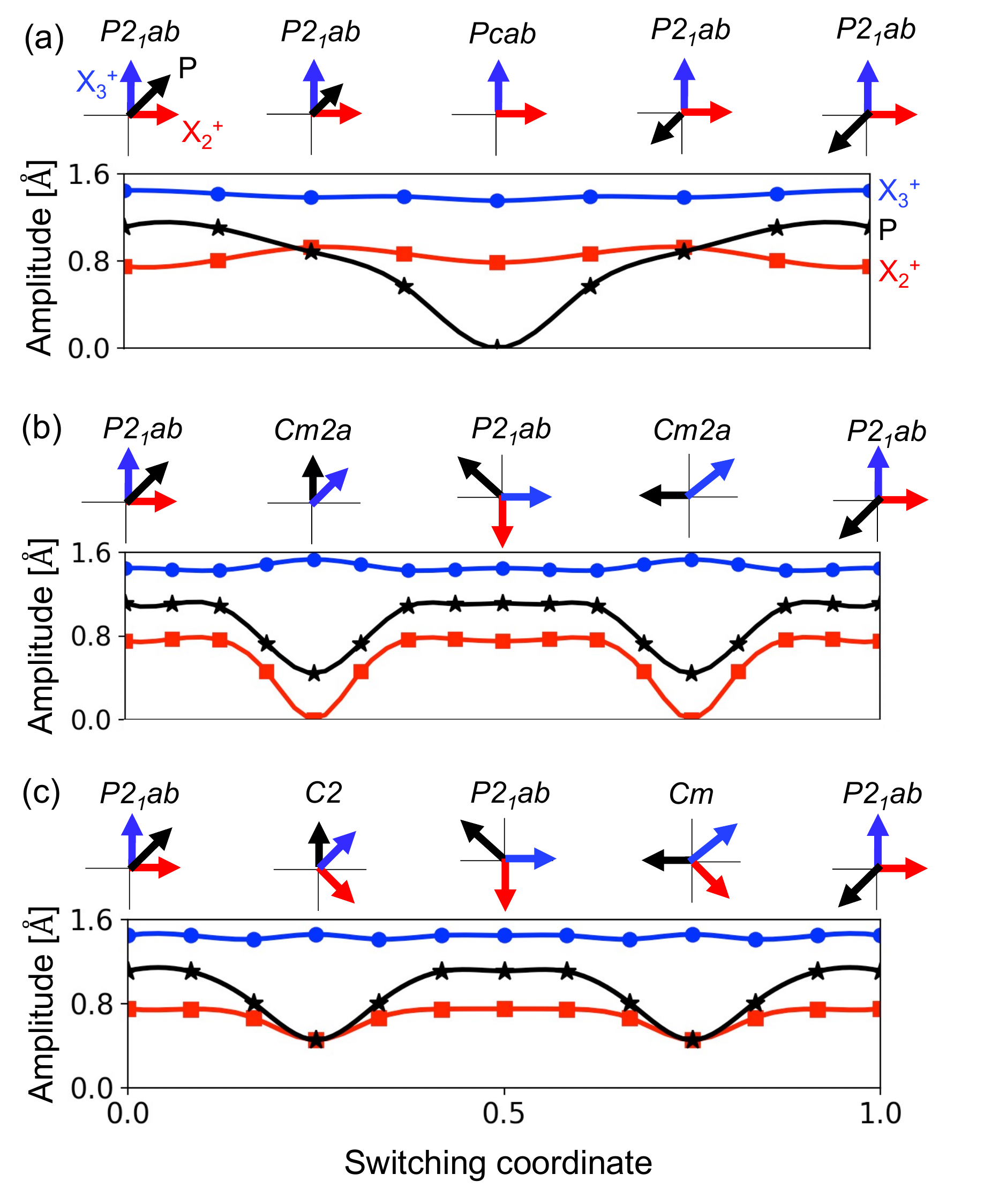}
 \captionsetup{width=0.95\linewidth,justification=raggedright}
 \caption{Amplitudes of the $X_3^+$, $X_2^+$, and $\Gamma_5^-$ (\textbf{\textit{P}}) structural distortions as a function of switching coordinate for each of the switching paths shown in Figure~\ref{P_switching_pathways}: (a) one-step switching with barrier $Pcab$, (b) two-step switching with barrier $Cm2a$, and (c) two-step switching with barriers $C2/Cm$. The amplitudes are obtained from NEB calculations and are reported for a 36-atom computational cell. }
 \label{fig:mode_amplitudes}
\end{figure}

Figure~\ref{fig:mode_amplitudes} shows how the amplitudes of the $X_3^+$, $X_2^+$, and polar $\Gamma_5^-$  distortions evolve along each ferroelectric switching path, obtained from NEB calculations.
In the one-step switching path shown in Figure~\ref{fig:mode_amplitudes}(a),  the  polar distortion amplitude goes to zero at the  barrier structure $Pcab$, whereas the  $X_2^+$ and $X_3^+$ distortion amplitudes remain almost unchanged throughout the switching process. In the two-step switching path via $Cm2a$ shown in Figure~\ref{fig:mode_amplitudes}(b), at the $Cm2a$ barrier the polar distortion amplitude decreases by about half and that of $X_2^+$ amplitude goes to zero, whereas the   $X_3^+$ amplitude again changes very little throughout the switching process. Finally, for the two-step $C2/Cm$ path shown in Figure~\ref{fig:mode_amplitudes}(c), all three distortion amplitudes remain finite throughout the switching process, although the polar and $X_2^+$ amplitudes are suppressed upon approaching the $C2/Cm$ barriers. 

To summarize, we find that the two-step switching paths have lower energy barriers than the one-step switching path in Bi$_2$WO$_6$.  This implies that switching proceeds via two 90$^\circ$ steps, in agreement with the experimental observations of Ref.~\onlinecite{Wang2016d}. We find that the two-step paths that pass through $Cm/C2$ and $Cm2a$ have almost the same energy barrier ($\approx$100 meV/f.u.), with the barrier for the $Cm/C2$ path being slightly lower. We also investigate the epitaxial strain dependence of these energy barriers (Appendix~\ref{appendix2}), and find that the two-step barriers remain the lowest energy except possibly under highly compressive strains. The two-dimensional structural order parameters facilitate the lower energy two-step switching, which involves order parameter rotation rather than completely turning the polarization off/on. We make use of these ferroelectric switching paths in the next section to guide us to the relevant structural phases in which to explore the spin orientation of magnetic dopants in Bi$_2$WO$_6$.

\subsection{Spin directionality of magnetic dopants}

\begin{figure*}[htb]
 \centering
 \includegraphics[width=0.9\textwidth]{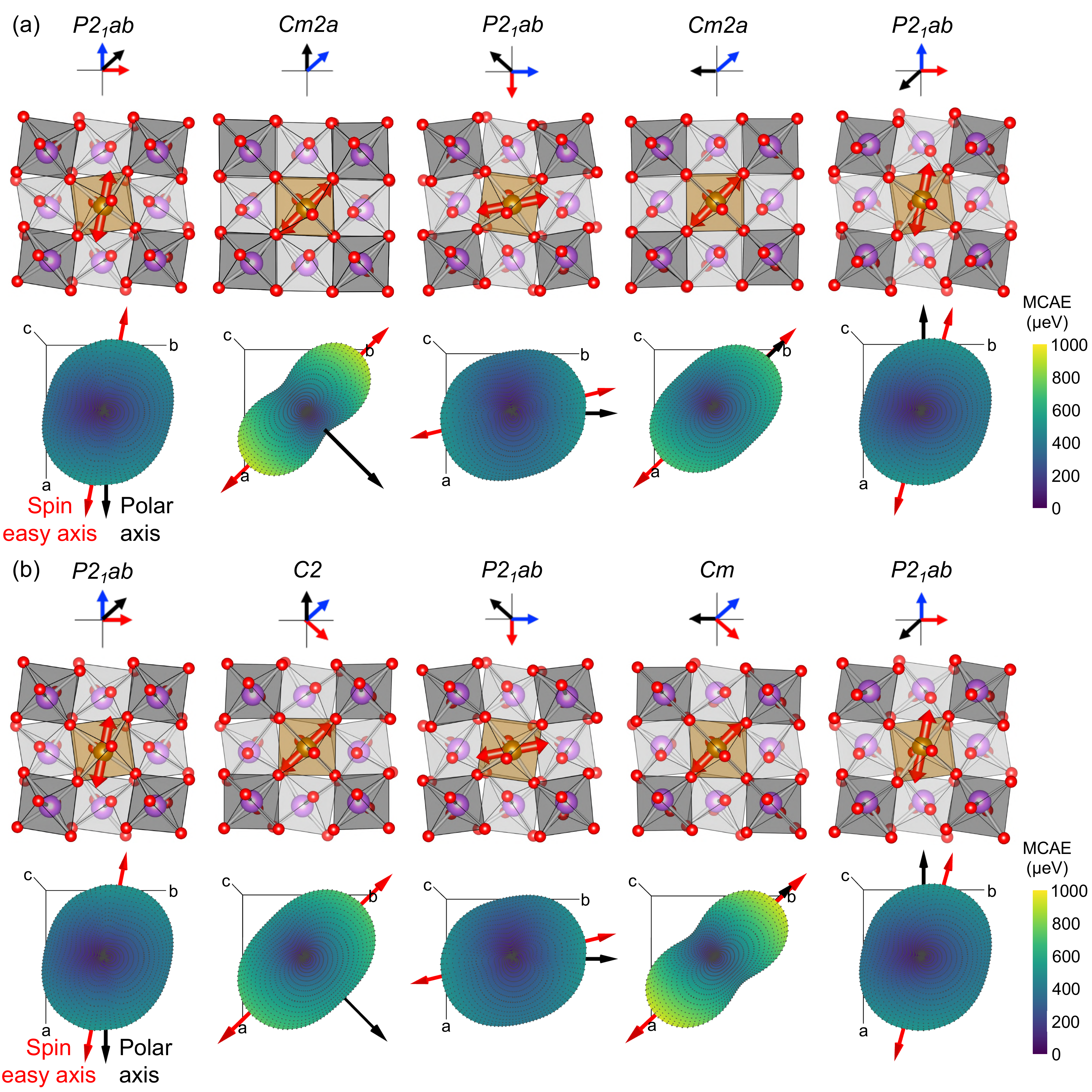}
 \caption{Change in spin directionality on an Fe-dopant during switching (a) via the $Cm2a$ intermediate phase and (b) via the $C2$ and $Cm$ phases. Magnetocrystalline anisotropy energy surfaces are plotted with red arrows indicating the spin easy directions and black arrows showing the polar axis in the $P2_1ab$ phases. In the crystal structures, purple, red and gold spheres are W, O and Fe respectively, and only \ce{WO6} octahedra are shown; the Fe-dopant sits in the upper layer with the dark grey octahedra and the lower layer is light grey. }
 \label{fig:spinpath}
\end{figure*}

\noindent
We next investigate the impact of polarization reversal on the spin orientation of magnetic dopants in Bi$_2$WO$_6$. 
Magnetic dopants in \ce{Bi2WO6} exhibit magnetocrystalline anisotropy -- a preferred directionality of the unpaired electron spins -- that arises due to the crystal field at the dopant site and spin-orbit coupling. Here, we consider Fe$^{3+}$ substitutional defects on W-sites, which is one of the potential defect species in \ce{Bi2WO6}. Although the $P2_1ab$ structure only contains one distinct W site, the barrier structures encountered in the two-step switching processes contain multiple symmetry-distinct W sites (and hence dopant positions), which are shown in Figure~\ref{fig:intermediate_structures}(b-d).  The energetics of this defect and alternative Fe-sites will be fully considered in a forthcoming work. 

Figure~\ref{fig:spinpath} tracks the change in the MCAE surface along the two lowest energy switching pathways, $P2_1ab$ $\rightarrow$ $C2$ $\rightarrow$ $P2_1ab$ $\rightarrow$ $Cm$ $\rightarrow$ $P2_1ab$ and $P2_1ab$ $\rightarrow$ $Cm2a$ $\rightarrow$ $P2_1ab$ $\rightarrow$ $Cm2a$ $\rightarrow$ $P2_1ab$, revealing the change in directionality of the Fe-dopant spins during switching. Beginning in the $P2_1ab$ phase with the polarization \textbf{\textit{P}} oriented along [1~0~0], we identify a spin easy axis that lies in the crystallographic $ab$-plane at 11$\degree$ with respect to \textbf{\textit{P}}. In the octahedral orientation indicated in the first step, the spin easy axis is along $\langle1.0,~-0.2,~0.0\rangle$. The calculated magnitude of the MCAE is 530~$\mu$eV, which is the energy difference between the $x$ and $z$ principal axes of the MCAE surface.
In addition, we calculate an in-plane anisotropy between the $x$ and $y$ principal axes of 130~$\mu$eV. The principal axes of the MCAE surface for each of the switching steps are given in Appendix~\ref{appendix3}.   

Taking the $Cm2a$ path (Figure~\ref{fig:spinpath}a), the Fe-dopant (along with all the octahedra in the top layer) first passes through the 4$c$ Wyckoff position in $Cm2a$, which has site symmetry $m$. On this site the MCAE is significantly increased to 940~$\mu$eV and the spin easy axis is $\langle-0.7,~0.7,~0.0\rangle$. At 90$\degree$ switching, the structure returns to the $P2_1ab$ phase with spin easy axis now along $\langle-0.2,~1.0,~0.0\rangle$, a rotation of 112$\degree$ around the $c$-axis from the original spin easy axis. Passing through the $Cm2a$ structure a second time, the dopant (and all octahedra in the top layer) is in the 4$b$ Wyckoff position with site symmetry $2$, resulting in a slightly lower MCAE of 730~$\mu$eV compared to the 4$c$ site, although the spin easy axis remains the same. This can be understood by the variation in off-centering of Fe, making a 170$\degree$ O--Fe--O bond angle in the 4$c$ position compared to 180$\degree$ in the 4$b$ position (taking the bonds aligned parallel to the spin axis). In the last step, the structure returns to $P2_1ab$ with \textbf{\textit{P}} switched by 180$\degree$ and the spin easy axis returned to $\langle1.0,~-0.2,~0.0\rangle$. (Beginning in one of the orientations in the lower layer, the Fe-dopant would pass through the 4$b$ Wyckoff position first and the 4$c$ position second.)

Alternatively, taking the $C2/Cm$ path (Figure~\ref{fig:spinpath}b), the Fe-dopant first passes through the $C2$ phase in Wyckoff position 2$b$ (site symmetry $2$) and secondly through the $Cm$  phase in Wyckoff position 2$a$ (site symmetry $m$). The MCAE values are 750~$\mu$eV and 970~$\mu$eV respectively, and the spin easy axis is along $\langle-0.7,~0.7,~0.0\rangle$ in both cases. As in the $Cm2a$ path, a 90$^\circ$ switch in the polarization direction results in a rotation of the spin easy axis by 112$^\circ$ about the $c$-axis. The similarity in crystal field environment in the two switching pathways accounts for the resemblance between the MCAE surfaces in Figure~\ref{fig:spinpath}a-b.
Details of the site symmetries and in-plane and out-of-plane MCAE are listed in Table~\ref{tab:mcae}, including alternative dopant positions. There are symmetrically inequivalent W-sites on the 2$a$ Wyckoff position in $Cm$ and the 2$b$ position in $C2$, which have a slightly different crystal field environment from the sites considered above, resulting in small differences in the MCAE values. In addition, an alternative domain choice for the switching pathways could have taken the Fe-dopant through the 4$b$ position in $Cm$ and the 4$c$ position in $C2$. MCAE data is not available for the latter as the structural optimization on this site did not converge the forces on the ions below a reasonable number.

The MCAEs reported here are typical of 3$d$ transition metal atoms,~\cite{schron2012crystalline,rau2014reaching} and their low magnitudes indicate that thermally induced switching could occur if not kept at very low temperatures. This could be sufficient for devices operating at cryogenic temperatures (e.g. quantum computing); an MCAE of 0.5~meV corresponds to an energy barrier of  $\sim$6~K, three orders of magnitude greater than the typical mK operating temperatures of many quantum devices. However, strategies to increase the MCAE should be explored to exclude thermally induced switching in higher temperature applications. These could include systems with reduced dimensionalities and 4$d$ or 5$d$ transition metal and rare earth atoms which have been shown to exhibit giant MCAE values.\cite{jiexiang2018,levzaic2011high,rout2021large,zhou2015giant}


\begin{table}[]
\caption{\label{tab:mcae} Fe$^{3+}$ dopant positions, site symmetries and magnetocrystalline anisotropy energies (MCAE) in $P2_1ab$ and intermediate switching phases. * indicates the site chosen for the switching pathway in Figure~\ref{fig:spinpath}b, where multiple sites are available.}
\begin{tabular}{lccccccc}
\hline
\hline
Phase    & Wyckoff & Site & \multicolumn{2}{c}{MCAE ($\mu$eV)} \\
         &    site        &   symmetry     & Out-of-plane     & In-plane \\
         \hline
$P2_1ab$   & 4$a$ & 1   &  530  &  130  \\
$Cm2a$     & 4$b$ & 2   &  730  &  210  \\
           & 4$c$ & $m$ &  940  &  510  \\
$Cm$       & 2$a$* & $m$ &  970  &  520  \\
           & 2$a$ & $m$ &  980  &  530  \\
           & 4$b$ & 1   &  650  &  130  \\
$C2$       & 2$b$* & 2   &  750  &  210  \\
           & 2$b$ & 2   &  760  &  200  \\
           & 4$c$ & 1   &  N/A  &       \\ 
                      \hline
                      \hline
\end{tabular}
\end{table}

\section*{Conclusion}
\noindent
We use a combination of group theoretic analysis and DFT calculations to determine the intrinsic  ferroelectric switching pathways of \ce{Bi2WO6}. We identify several pathways: a one-step pathway, via $Pcab$, and three two-step pathways, via $C2/Cm$, $Cm2a$ and $Cm2m$. By comparing energies of the barrier structures we find that the two-step paths are lower energy than the one-step path, in agreement with experiment:~\cite{Wang2016d} in particular,  the $C2/Cm$ path provides the lowest energy barrier of 97-99 meV/f.u. and the $Cm2a$ barrier is only slightly higher at 103 meV/f.u.. These intrinsic switching barrier energies are comparable to those of other structurally complex ferroelectrics such as LiNbO$_3$ and Ca$_3$Ti$_2$O$_7$, which have barriers of 130 meV/f.u.~\cite{Ye2016} and 64 meV/f.u.~\cite{nowadnick2016domains}, respectively.  

Magnetic defects experience a change in crystal field environment during switching, resulting in a change in magnetic anisotropy  at each switching step. Contrasting with Fe$^{3+}$ dopants in PbTiO$_3$ which exhibit a spin easy plane~\cite{Liu2021}, the lower crystallographic symmetry of Bi$_2$WO$_6$ results in a spin easy axis. By calculating MCAE surfaces, we find how the spin orientation of  a Fe$^{3+}$ substitutional defect on W-sites changes during polarization switching. In the $P2_1ab$ structure, the spin easy axis is in the $ab$-plane at 11$\degree$ with respect to the polar axis and has an out-of-plane MCAE of 530~$\mu$eV and an in-plane MCAE of 130~$\mu$eV. During ferroelectric switching via intermediate phases, the spin easy axis  rotates within the $ab$-plane and the MCAE is considerably increased (650-980~$\mu$eV out-of-plane, 130-530~$\mu$eV in-plane) due to changes in the local crystal environment of the dopant. 
We find that switching the polarization by 90$^\circ$ in Bi$_2$WO$_6$ results in a 112$^\circ$ rotation of the spin easy axis. However, a full 180$^\circ$ reversal of the polarization returns the spin easy axis to its original orientation. 

Based on these results, we suggest that a possible pathway to achieve full  spin control with 180$^\circ$ polarization switching is to consider ferroelectrics where an additional structural distortion that couples to the polarization must change during 180$^\circ$ switching. If this structural distortion also impacts the magnetic anisotropy, then the change to the distortion due to polarization switching may result in a different spin easy axis in the $+$\textbf{\textit{P}} and $-$\textbf{\textit{P}} states. For example, changes to octahedral rotation distortions, which are extremely common in (layered) perovskite oxides,  can modify spin easy planes and axes.~\cite{liao2016controlled,yi2017tuning} Although the $X_3^+$ and $X_2^+$ octahedral rotation patterns in Bi$_2$WO$_6$ change along the two-step switching paths, the octahedral rotation amplitudes and pattern in the starting ($+$\textbf{\textit{P}}) and final ($-$\textbf{\textit{P}}) structures are the same. However, there are other layered perovskite ferroelectrics, such as the Aurivillius compound SrBi$_2$Ta$_2$O$_9$~\cite{perez2004competing} and several $n$=2 Ruddlesden-Popper ferroelectric oxides,~\cite{benedek2011hybrid,nowadnick2016domains} where reversal of the polarization requires \textit{by symmetry} that the sense of an octahedral rotation also reverse. Compounds such as these may provide the necessary ingredients to create distinct low symmetry environments and hence different  MCAE surfaces in polarization reversed states.      


\section*{Acknowledgements}
\noindent
This work was supported by the Microelectronics Co-Design Research Program, under the Office of Science of the U.S. Department of Energy under Contract No. DE-AC02-05CH11231 (K.I., N.L., and S.M.G.). N.P., Z.C., S.P.R., and E.A.N. acknowledge support from University of California, Merced. Computational resources were provided by the National Energy Research Scientific Computing Center and the Molecular Foundry, DOE Office of Science User Facilities supported by the Office of Science, U.S. Department of Energy under Contract No. DE-AC02-05CH11231.  The work performed at the Molecular Foundry was supported by the Office of Science, Office of Basic Energy Sciences, of the U.S. Department of Energy under the same contract. We also acknowledge the use of computational resources supported by the Center for Functional Nanomaterials, which is a U.S. DOE Office of Science Facility, and the Scientific Data and Computing Center, a component of the Computational Science Initiative, at Brookhaven National Laboratory under Contract No. DE-SC0012704. In addition, this work used the Extreme Science and Engineering Discovery Environment (XSEDE) Expanse cluster at the San Diego Supercomputing Center through allocation TG-PHY200085.


\section*{Author contributions}
K.I. and N.P. contributed equally to this work.

\appendix

\setcounter{figure}{0}
\setcounter{table}{0}
\renewcommand{\thefigure}{A\arabic{figure}}
\renewcommand{\thetable}{A\arabic{table}}

\section{Additional subgroups of $I4/mmm$}
\label{appendix1}

Table~\ref{tab:irreps_details} reports the subgroups of $I4/mmm$ generated by distinct directions of the $\Gamma_5^-$, $X_2^+$, and $X_3^+$ order parameters, along with the distortion amplitudes, lattice parameters, and total energies of Bi$_2$WO$_6$ after DFT structural relaxations in each space group.

\begin{table*}[]
\caption{\label{tab:irreps_details} Subgroups of $I4/mmm$ established by distinct directions of the $\Gamma_5^-$, $X_2^+$, and $X_3^+$ order parameters. Total energies, distortion amplitudes, and lattice parameters obtained from DFT structural relaxations of Bi$_2$WO$_6$ in each space group are given. The energies are reported relative to the energy of $P2_1ab$, which is set to 0 meV/f.u., and the distortion amplitudes  are obtained by decomposing the distorted structures with respect to $I4/mmm$ and are reported for a 36-atom computational cell. }

\begin{tabular}{c  c  c  c  c c c c }
\hline
\hline
Irrep & Direction & Space group (N$^o$) & Amplitude ({\AA})& \multicolumn{3}{l}{Lattice parameters ({\AA}}) & Energy (meV/f.u.)  \\
       
& & & & $a$ & $b$ & $c$ & \\

\hline

$\Gamma_5^-$               &   $(a,0)$     & $Imm2$ (44) & 1.47  &  5.472 & 5.472 & 16.575 & 154.54\\
                                    &  $(a,a)$      & $Fmm2$ (42) & 1.25    &  5.504  & 5.461 & 16.436 & 133.91\\      
                                    \hline

$X_2^+$               &   $(a,0)$     & $Bbcm$ (64)  & 1.31  &  5.290 & 5.290 & 16.646 &                                      231.32\\
                                    &   $(a,a)$ &  $P4/mbm$ (127)  & 0.80   &  5.349  & 5.349 & 16.524 & 310.16\\      
                                    \hline

$X_3^+$               &    $(a,0)$ & $Bbcm$ (64)  & 1.48 &  5.368 & 5.395 & 16.663 & 178.40\\
                                     & $(a,a)$ & $P4_2/ncm$ (138)  &  1.51  &  5.393  & 5.393 & 16.576 & 161.40\\      
                                    \hline
\end{tabular}
\end{table*}


\section{Epitaxial strain}
\label{appendix2}
\begin{figure}[]
 \centering
 \includegraphics[width=0.45\textwidth]{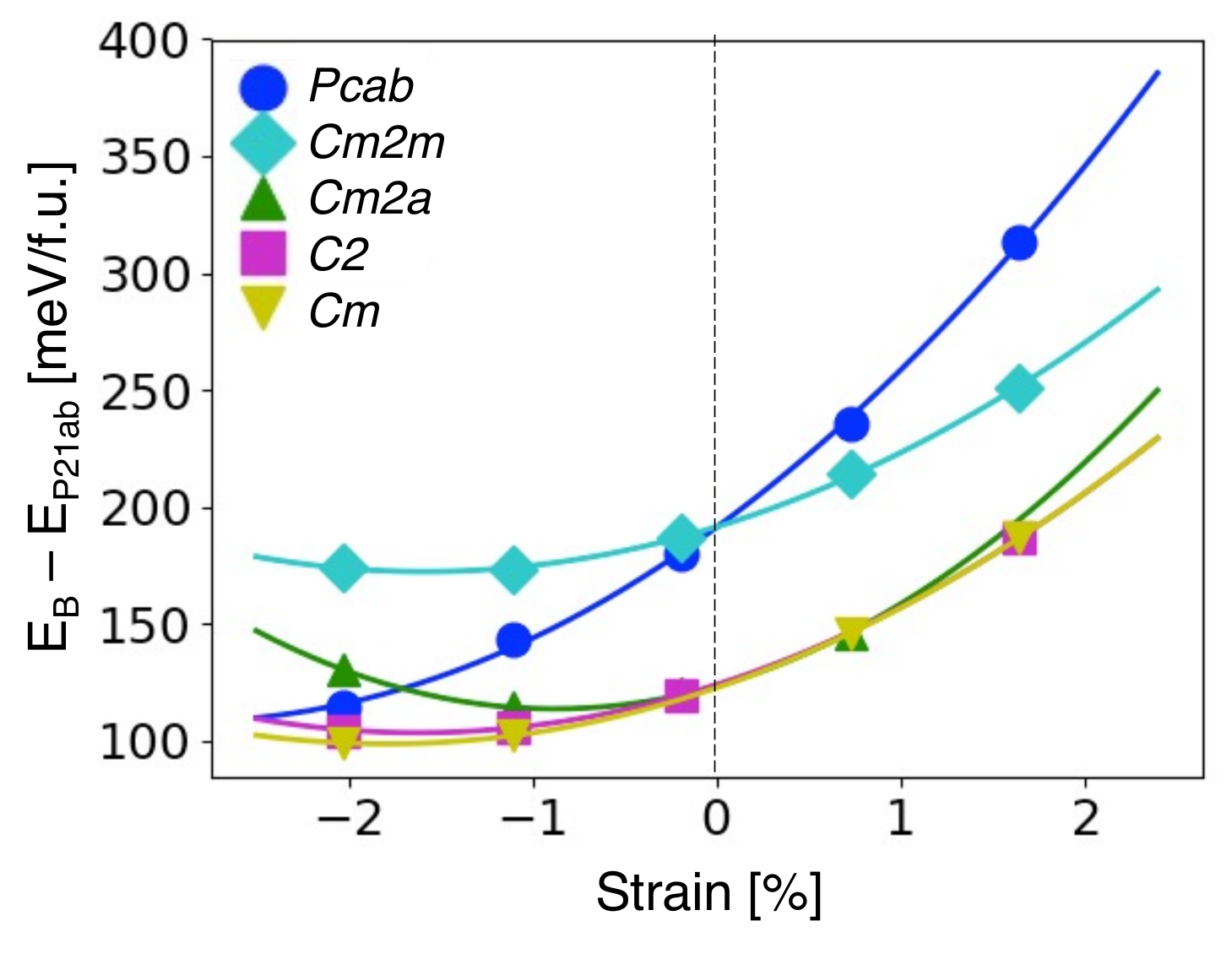}
 \caption{ Energy of the ferroelectric switching barrier structures $B$ above $P2_1ab$ as a function of epitaxial strain. The strain is applied biaxially in the $ab$ plane, and 0\% strain is defined with respect to the $P2_1ab$ lattice parameters.}
 \label{fig:strain}
\end{figure}

Figure~\ref{fig:strain} presents the energies of the ferroelectric switching barrier structures  discussed in the main text as a function of epitaxial strain. In these calculations, biaxial strain is applied in the $ab$-plane, and the lattice parameter $c$ and all atomic positions are allowed to relax. 
We find that all energy barriers increase upon going from compressive to tensile strain. The $Pcab$ energy barrier changes the most dramatically with strain. The $C2/Cm$ structures provide the lowest energy barriers at all strains that we consider, although under highly compressive strains (larger than 2$\%$) the $Pcab$ barrier may become lowest, which would suggest a crossover from two- to one-step switching being the lowest energy path. 

\section{Principal axes of magnetocrystalline anisotropy energy surfaces}
\label{appendix3}

Table~\ref{tab:principal_axes} reports the principal axes of the MCAE surfaces  shown in Figure~\ref{fig:spinpath}. These axes are orthonormal vectors which describe the three principal axes of rotation of the MCAE surface. The spin easy axis lies along the $x$ principal axis.

\begin{table*}[]
\caption{\label{tab:principal_axes} Principal axes of the magnetocrystalline anisotropy energy surfaces for each of the switching steps shown in Figure~\ref{fig:spinpath}.}
\begin{tabular}{lcccccc}
\hline
\hline
Switching & Phase & Wyckoff & \multicolumn{3}{c}{Principal axes} \\
figure    &       & site    & $x$ & $y$ & $z$ \\
\hline
~\ref{fig:spinpath}(a) & $P2_1ab$  & 4$a$ & [1.0~-0.2~0.0] & [0.2~0.9~0.3] & [-0.1~-0.3~0.9] \\
                       & $Cm2a$    & 4$c$ & [0.7~-0.7~0.0] & [0.6~0.6~-0.5] & [-0.3~-0.3~-0.9]  \\
                       & $P2_1ab$  & 4$a$ & [-0.2~1.0~0.0] & [-0.9~-0.2~-0.3] & [-0.3~-0.1~0.9] \\
                       & $Cm2a$    & 4$b$ & [0.7~-0.7~0.0] & [-0.7~-0.7~-0.3] & [0.2~0.2~-0.9] \\
                       & $P2_1ab$  & 4$a$ & [-1.0~0.2~0.0] & [0.2~0.9~0.3] & [0.1~0.3~-0.9] \\
\hline
~\ref{fig:spinpath}(b) & $P2_1ab$  & 4$a$ & [1.0~-0.2~0.0] & [0.2~0.9~0.3] & [-0.1~-0.3~0.9] \\
                       & $C2$      & 2$b$ & [0.7~-0.7~0.0] & [-0.7~-0.7~-0.3] & [0.2~0.2~-1.0] \\
                       & $P2_1ab$  & 4$a$ & [-0.2~1.0~0.0] & [-0.9~-0.2~-0.3] & [-0.3~-0.1~0.9] \\
                       & $Cm$      & 2$a$ & [-0.7~0.7~0.0] & [-0.6~-0.6~-0.4] & [-0.3~-0.3~0.9] \\
                       & $P2_1ab$  & 4$a$ & [-1.0~0.2~0.0] & [0.2~0.9~0.3] & [0.1~0.3~-0.9] \\
\hline
\hline
\end{tabular}
\end{table*}

\bibliography{refs}
\end{document}